\documentclass[journal]{IEEEtran}
\usepackage{verbatim}
\usepackage{amssymb}
\usepackage{diagbox}
\usepackage{cite, setspace,graphicx,graphics, subfigure,subfloat,amsmath,amssymb,amsfonts,stfloats,multirow,booktabs, url}
\usepackage[table,xcdraw]{xcolor}
\usepackage[below]{placeins}
\DeclareGraphicsExtensions{.pdf,.jpeg,.png}

\ifCLASSINFOpdf

\else

\fi

\begin{document}

%
\title{Reversible Data Hiding in Encrypted Images based on Pixel Prediction and Bit-plane Compression}
%
%
%
\author{Zhaoxia~Yin,~\IEEEmembership{Member,~IEEE,}
        Yinyin~Peng,~Youzhi~Xiang
\thanks{This research work is partly supported by National Natural Science Foundation of China (61872003, U1636206,61860206004).}
\thanks{Zhaoxia Yin,Yinyin~Peng and Youzhi Xiang are with the school of Key Laboratory of Intelligent Computing and Signal Processing, Ministry of Education, Anhui University, Hefei 230601, P. R. China, e-mail: yinzhaoxia@ahu.edu.cn.}}

\markboth{Journal of \LaTeX\ Class Files,~Vol.~14, No.~8, August~2015}%
{Shell \MakeLowercase{\textit{et al.}}: Bare Demo of IEEEtran.cls for IEEE Journals}
%



\maketitle

\begin{abstract}
Reversible data hiding in encrypted images (RDHEI) receives growing attention because it protects the content of the original image while the embedded data can be accurately extracted and the original image can be reconstructed lossless. To make full use of the correlation of the adjacent pixels, this paper proposes an RDHEI scheme based on pixel prediction and bit-plane compression.
Firstly, to vacate room for data embedding, the prediction error of the original image is calculated and used for bit-plane rearrangement and compression. Then, the image after vacating room is encrypted by a stream cipher. Finally, the additional data is embedded in the vacated room by multi-LSB substitution. Experimental results show that the embedding capacity of the proposed method outperforms the state-of-the-art methods.
\end{abstract}


%
\IEEEpeerreviewmaketitle

\section{Introduction}
%
%
%
%
\IEEEPARstart{R}{eversible} data hiding (RDH) has attracted wide attention that it embeds additional data into multimedia. At the same time, the original medium can be reconstructed lossless and the embedded data can be accurately extracted. Therefore, RDH is widely utilized in many applications [1], such as reversible image processing, reversible visual transformation and reversible adversarial example described, etc. Many RDH algorithms in spatial images have been proposed and can be classified into three categories based on the different technologies: lossless compression [2][3], histogram shifting [4]-[7], and difference expansion [8][9]. Recently, with the widely used cloud storage and privacy protection, multimedia encryption may be implemented before transmission for the content-owner. Thus, reversible data hiding in encrypted images (RDHEI) [10][11] has been proposed.
\par Generally, content-owner, data-hider, and receiver are three end-users of RDHEI [12]. The contents of the original image are protected by the encryption operation of the content-owner. The data-hider can embed additional data without knowing the contents of the original image. The permissions depend on the number of keys for the receiver. The data encryption key can accurately decrypt the extracted additional data, and the image encryption key can reconstruct the original image.
\par The RDHEI methods proposed in recent years mainly can be classified into two categories: vacating room after encryption (VRAE) [12]-[15] and reserving room before encryption (RRBE) [16]-[22], as shown in Fig. 1. For the VRAE methods, the local spatial correlation of the original image is utilized while the image is encrypted. But they can't achieve satisfied embedding capacity and some algorithms are irreversible or inseparable. The RRBE methods reserve room before image encryption based on the pixel correlation. Then, the reserving room is transferred into the encrypted image. Compared with the VRAE method, the RRBE algorithm makes full use of the pixel correlation of the original image and has a higher embedding capacity.
\begin{figure*}[!ht]
\centering
\subfigure[]{\includegraphics[width=5in]{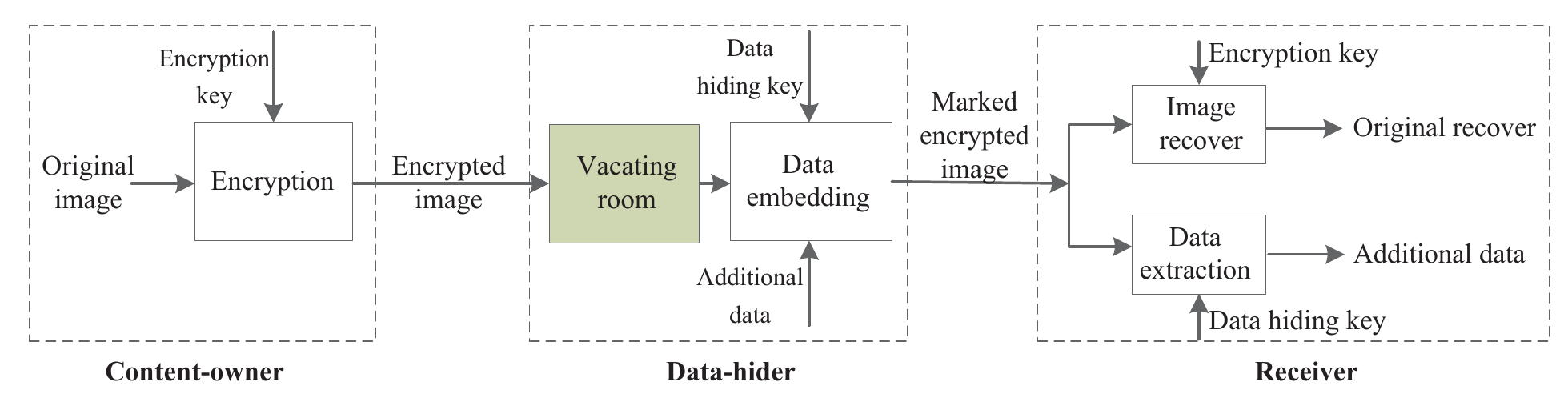}%
\label{fig-1-a}}
\hfil
\subfigure[]{\includegraphics[width=5in]{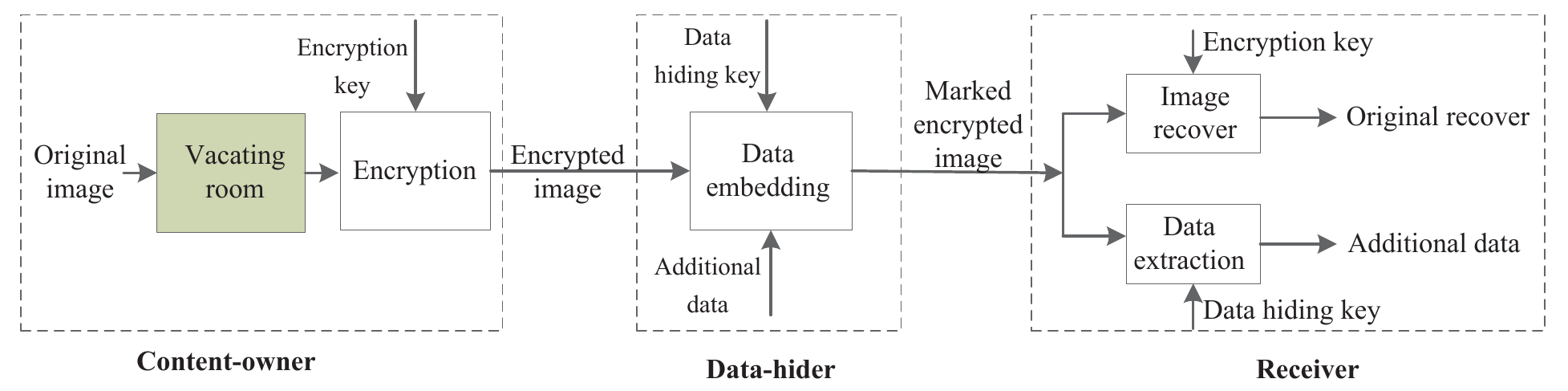}%
\label{fig-1-b}}
\caption{Two different frameworks of RDHEI methods: (a) VRAE and (b) RRBE.}
\label{fig-1}
\end{figure*}

\par In [16], The RRBE scheme is first proposed. The least significant bits (LSBs) of some pixels are embedded into other pixels before encryption. The embedding capacity has great improvement compared with the traditional VRAE methods [11]-[13]. At the same time, additional data can be extracted accurately and images can be reconstructed. In [17],  an estimation technique is proposed to reserve room before encryption. In order to further improve the embedding capacity of RDHEI, Chen ${ et~al. }$ [18] proposed an RDHEI method based on bit-plane compression. Firstly, the bit-plane rearrangement (BPR) is implemented. Then, the rearranged bit-stream is compressed. Finally, the vacated room is used to embed additional data. Different from the bit-plane compression, several methods based on image marking [19]-[22] are also proposed. Puteaux ${ et~al. }$ [19] proposed an RDHEI method based on the most significant bit (MSB) prediction. Firstly, the error label map recorded the non-embeddable pixels is embedded into the encrypted image. Then, the MSB of the available pixels is substituted by 1 bit additional data according to the error label map. Based on this method, an RDHEI of two-MSB prediction scheme is proposed in [20]. A parametric binary tree labeling method presented in [21] is used to distinguish the prediction errors. The data-hider can use the parameter to embed the additional data in the encrypted image. In another method proposed by Yin ${ et~al. }$ [22], the binary sequence of the original pixel and the prediction error is sequentially compared from the MSB to the LSB. Then, the number of same bits is marked with Huffman coding. Finally, the multi-MSB is substituted by the additional data and the embedding capacity increases. Compared with the above method, the embedding capacity has increased. But the correlation of the adjacent pixels isn't fully utilized and embedding capacity can be further increased.
\par How to make full use of the correlation of the adjacent pixels? In this paper, an RDHEI scheme based on pixel prediction and bit-plane compression is proposed. At the content-owner end, the prediction error of the original image is first calculated. Then, the BPR and the bit-stream compression (BSC) algorithms are implemented. The vacated room is used to embed additional data is embedded by the data-hider in the vacated room. Finally, the receiver can independently reconstruct the image or extract additional data according to the different keys. The bit-plane of the prediction error is compressed in the proposed method. Each bit of all pixels is fully utilized. Experimental results show that the proposed method outperforms the state-of-the-art methods.
\par The remainder of this paper is organized as follows: bit-plane rearrangement and compression schemes [18] are briefly described in Section II. An RDHEI scheme based on pixel prediction and bit-plane compression is detailed introduced in Section III. Experimental results and the comparison are given in Section IV and Section V concludes this paper.

\section{Bit-Plane Rearrangement and Compression}
In this chapter, the bit-plane rearrangement and the bit-stream compression algorithms [18] are introduced. In order to make full use of the spatial correlation in the original images, the bit-planes divided into blocks are rearranged using four types. Section II-A gives detailed steps and example. Section II-B also gives detailed steps and example to introduce the bit-stream compression algorithm.
\begin{figure*}[!ht]
  \centering
    \includegraphics[width=4in]{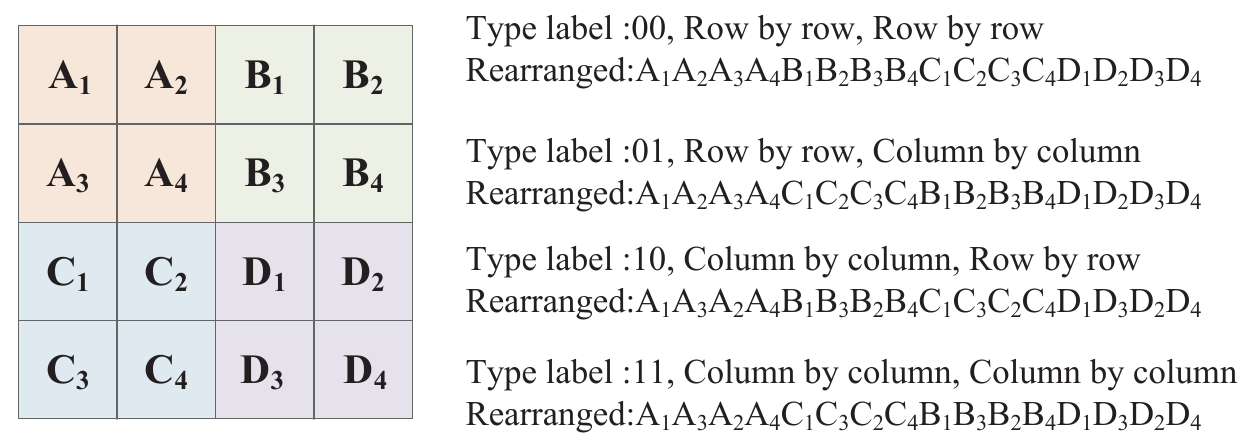}
  \caption{Four types of rearrangement.}
\label{fig-2}
\end{figure*}

\begin{figure*}[!ht]
  \centering
    \includegraphics[width=4in]{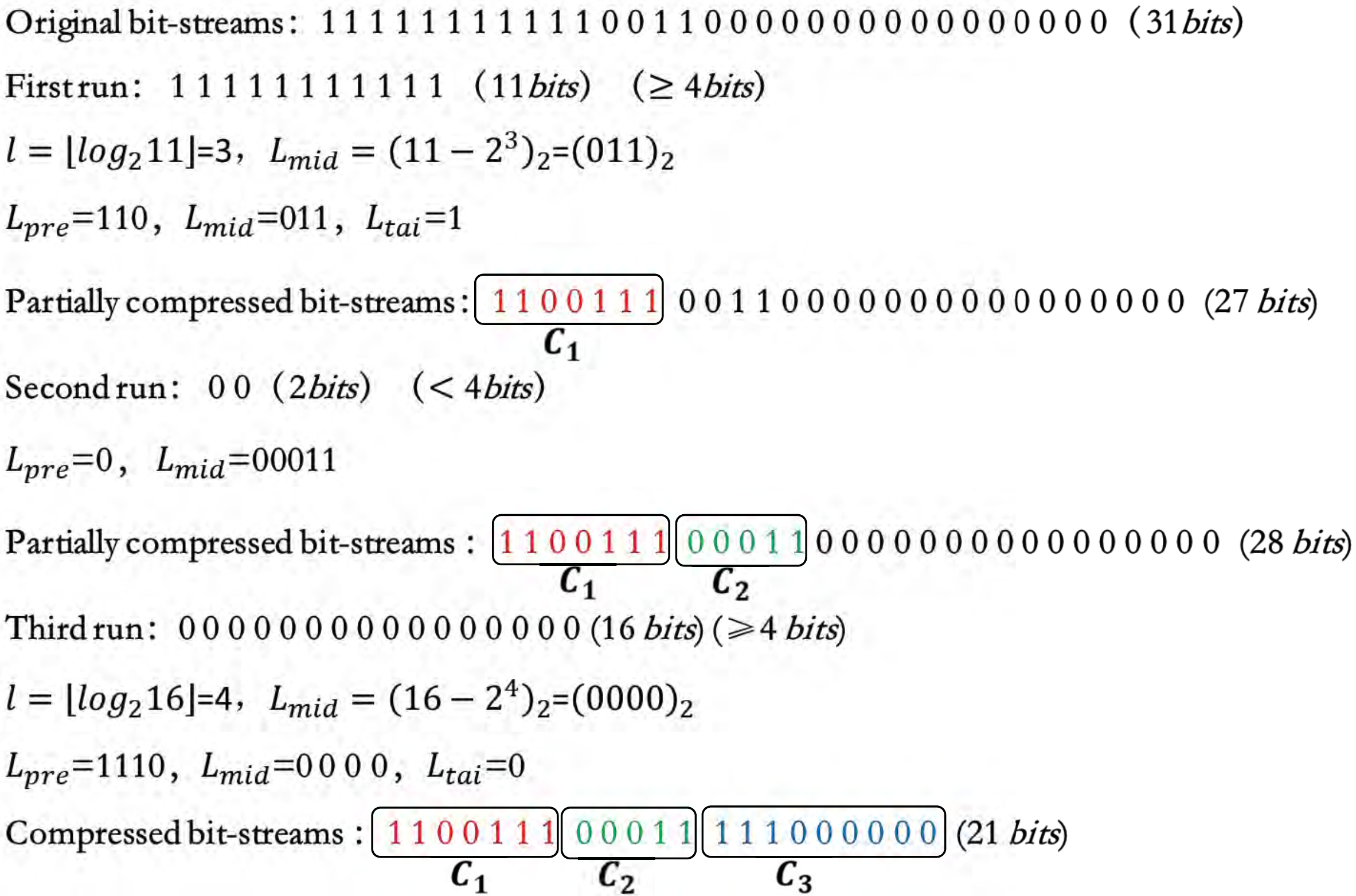}
  \caption{Example of the BSC procedure.}
\label{fig-3}
\end{figure*}
\subsection{Bit-Plane Rearrangement}
As we all know, the size of a natural image is relatively large. If each bit-plane of the image is traversed by raster scanning, the spatial correlation of the pixels is not fully utilized. Thus, in order to improve the compression rate, the bit-streams of the bit-plane are rearranged by block scanning in [18]. Firstly, the bit-planes are divided into blocks of the same size. Then, the bit-planes are rearranged by the four types. Taking Fig. 2 as an example to introduce the main step. The bit-plane is first divided into $ 2\times 2 $ blocks. Type label = 00 denotes that the traversing order of bits inside the block is row by row, and the order of blocks is also row by row. The rearranged bit-stream sequence is $ A_{1}A_{2}A_{3}A_{4}B_{1}B_{2}B_{3}B_{4}C_{1}C_{2}C_{3}C_{4}D_{1}D_{2}D_{3}D_{4} $. Type label = 01 represents that the traversing order of bits inside the block is row by row, and the order of blocks is column by column. The rearranged bit-stream sequence is $ A_{1}A_{2}A_{3}A_{4}C_{1}C_{2}C_{3}C_{4}B_{1}B_{2}B_{3}B_{4}D_{1}D_{2}D_{3}D_{4} $. The rearranged results of the other two types can also be obtained from the Fig. 2.

\subsection{Bit-Stream Compression}
After using the BPR algorithm, the rearranged bit-streams will have a large number of adjacent ``0" or ``1". Due to this characteristic, the BSC algorithm vacates more room to embedding additional data. Firstly, extracting the same bits sequence $L$ from the rearranged bit-streams. Secondly, encoding $L$ as a code word. Finally, the above operation is repeatedly performed for the subsequent sequence of rearranged bit-streams. The number of $ L$ determines the compression mode:
\par Case 1: $ L< 4$. $L_{pre}= 0$, the value of $L_{mid} $ is the rearranged bit-streams of length $L_{fix} $ that is intercepted backward from the current bit.
\par Case 2: $L\geq 4$. $L_{pre}$ consists of $l-1$ sequential $1_{s}$ and end with 0. $L_{tai}=0$ or 1, indicating the repeat bits. The values of $l$ and $L_{mid}$ are calculated as follows:
\begin{equation}
l= \left \lfloor log_{2}L \right \rfloor
\end{equation}
\begin{equation}
L_{mid}=\left ( L-2^{l} \right )_{}2
\end{equation}
Where $L_{pre}$,  $L_{mid}$ and $L_{tai}$ denote prefix, middle and tail of $C_{1}$. $L_{fix}$ is a parameter and $\left \lfloor \ast  \right \rfloor$ is the floor operation.
\par For example, as shown in Fig. 3, assume the uncompressed bit-streams is 1 1 1 1 1 1 1 1 1 1 1 0 0 1 1 0 0 0 0 0 0 0 0 0 0 0 0 0 0 0 0 and $L_{fix}=4$ . In the first run process, firstly, the eleven $1_{s}$ is extracted. Then, 11 bits $\geq $ 4 bits, $l=3$ can be calculated by Eq. (1) and $L_{pre}=110$. $L_{mid}=011$ can be calculated by Eq. (2). Since the values of these 11 bits are all 1, $L_{tai}=1$. Finally, $C_{1}$ is 1 1 0 0 1 1 1. The second run of two $0_{s}$ is extracted. Because of 2 bits $<$ 4 bits, $L_{pre}=0$ and $L_{mid}=00011$. The third run of sixteen  $0_{s}$ is encoded as 111000000. Finally, the 31 original bits are compressed into 21 bits, and the vacated room can be used to embed additional data.

\section{Proposed Method}
This paper proposes an RDHEI based on pixel prediction and bit-plane compression. As shown in Fig. 4, three stages consist of our method: 1) Vacating room and image encryption by the content-owner; 2) Data embedding by the data-hider; 3) Data extraction and image decryption by the receiver. In the first stage, to vacate more room, the prediction error of the original image is used for bit-plane rearrangement and compression in the proposed method. Section III-A gives a detailed introduction. Then, the image encryption is performed in Section III-B. In the second stage, the additional data is embedded in the vacated room that is generated in the first stage. For the receiver, the image can be recovered and the embedded data can be extracted, separately.
\begin{figure*}[!ht]
  \centering
    \includegraphics[width=5in]{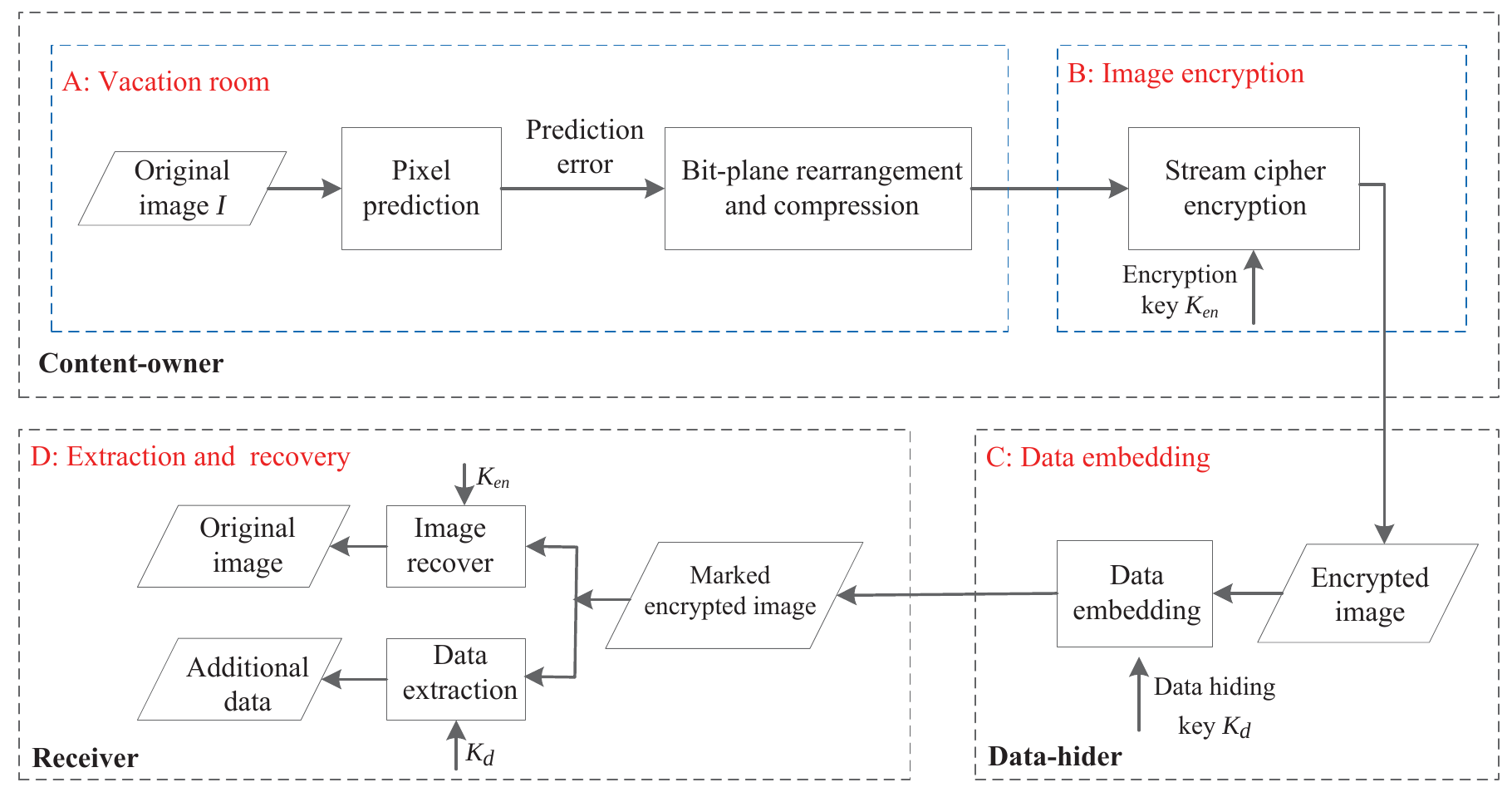}
  \caption{The framework of the proposed method.}
\label{fig-4}
\end{figure*}

\subsection{Vacating Room}
\subsubsection{Pixel Prediction}
Consider an original image $I$ of size $m\times n$, the predicted value of each pixel can be calculated by the median edge detector  (MED) [23]. As shown in Fig. 5, the predicted value $px$ of current pixel $x$ is calculated by its three neighboring pixels $x_{1}$,$x_{2}$ and $x_{3}$. The predicted value and the predicted error of each pixel can be giving in the following:

\begin{figure}[!ht]
  \centering
    \includegraphics[width=0.10\textwidth]{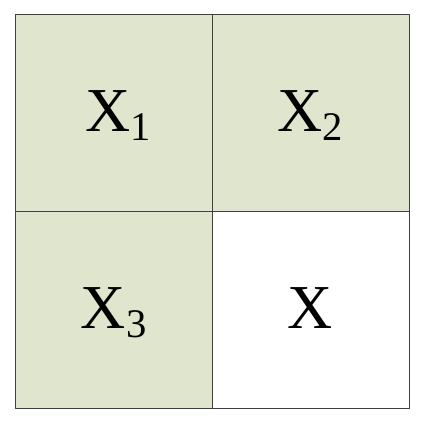}
  \caption{The prediction error of the current pixel by MED predictor.}
\label{fig-5}
\end{figure}

\begin{equation}
px= \begin{cases}
  max\left ( x_{2},x_{3} \right ), x_{1}\leq min\left ( x_{2},x_{3} \right )\\
  min\left ( x_{2},x_{3} \right ), x_{1}\geq  max\left ( x_{2},x_{3} \right ) \\
  x_{2}+x_{3}-x_{1},otherwise
\end{cases}
\end{equation}

\begin{equation}
ex=x-px
\end{equation}
\par In this part, the decimal prediction error $ex$ of the original image $I$ is converted into 8 bits binary. Because the sign of the prediction error can be positive or negative, the maximum significant bit (MSB) is used as the sign marker bit, ``1" denotes negative and ``0" denotes positive. The absolute value of predictive error is converted into 7 bits binary by the Eq. (5). When the prediction error exceeds the range of $\left [-127,127 \right ]$, this pixel is marked as the overflow pixel and does not be changed. At the same time, its location is recorded as auxiliary information.
\begin{equation}
ex^{k}\left ( i,j \right )=\left \lfloor \frac{ex\left ( i,j \right )}{2^{k-1}} \right \rfloor mod \:  \: 2,\, k=1,2...,7
\end{equation}
Where $ 1\leq i\leq m $ and $1\leq j\leq n$, $ex^{k}\left ( i,j \right )$ is the binary of corresponding bit. Taking the prediction error $ex\left ( 1,1 \right )=-2$ as an example, its binary of the absolute value is 0000010, so $ex^{8}\left ( 1,1 \right )=1$,  $ex^{2}\left ( 1,1 \right )=1$ and other bit-plane values of the corresponding bit are all 0.
\subsubsection{Bit-plane Rearrangement and Compression}
The BPR and the BSC algorithms are used for the 8 bits planes of the prediction error from the MSB plane to the LSB plane. Firstly, the MSB plane is rearranged using four types. Then, the shortest compressed bit-stream is selected from the four rearranged ways. At last, if the length of the compressed bit-stream is longer than that of the original bit-stream, the bit-plane is not compressed. If the length of the compressed bit-streams is less than that of the original bit-streams, the shortest compressed bit-streams is adopted.
\par As shown in Fig. 6, we take a bit-plane and type label = 00, 01 as an example. The length of the original bit-streams is 36 bits. Firstly, the bit-plane is divided into the block size of $3\times 3$. When type label = 00, the traversing order of bits inside the block and inter-block are all row by row, and the rearrangement result is 0 0 0 0 0 0 0 0 0 0 0 0 0 0 1 0 0 0 1 0 0 0 0 0 0 0 0 0 0 0 0 0 0 0 0 0. The compressed length is 24 bits. When type label = 01, the traversing order of bits inside the block is row by row and inter-block is column by column. The compressed length is 31bits. Since 24 bits $<$ 31 bits, the type label = 00 is adopted.
\begin{figure*}[!ht]
  \centering
    \includegraphics[width=5in]{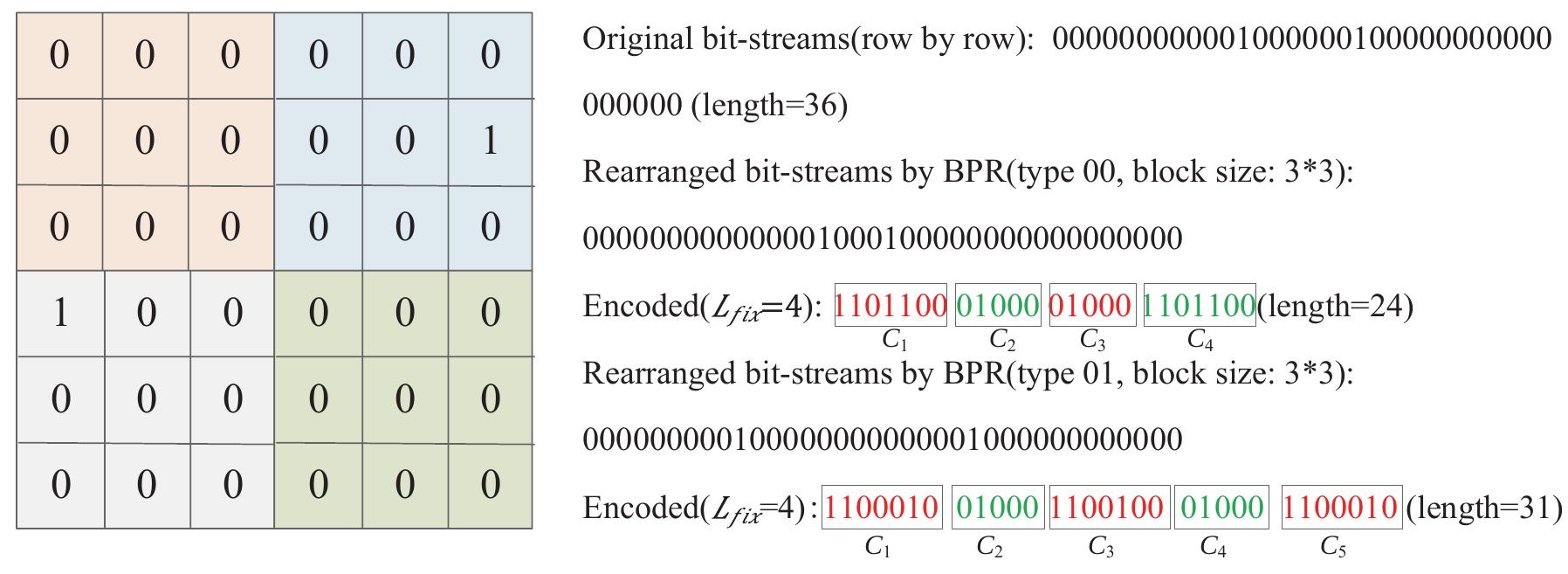}
  \caption{Example of the BPR and the BSC algorithms procedure.}
\label{fig-6}
\end{figure*}
\par For a compressed bit-plane, as shown in Fig. 7, the flag denoted whether the bit-plane can be compressed and is marked by 1 bit. The rearrangement type use 2 bits to record. The gray region represents the compressed bit-streams, and the blank region denotes the vacated room. Then, all compressed bit-streams and uncompressed bit-plane are connected to generate new multi-MSB planes and all the vacated multi-LSB planes are complemented by 0. Auxiliary information includes the block size, the value of $L_{fix}$, and the number of overflow pixels. As shown in Fig. 8, auxiliary information is placed on the MSB plane, the number of compressed bit-streams is placed on the LSB plane. Finally, the compressed image containing embeddable room $I_{c}$  is calculated by Eq. (6).
\begin{equation}
x_{c}\left ( i,j \right )=\sum_{k=1}^{8}x_{c}^{k}\left ( i,j \right )\times 2^{8-k},\: k=1,2...,8
\end{equation}
Where $1\leq i\leq m$ and $1\leq j\leq n$, $x_{c}\left ( i,j \right )$ is the pixel of compressed image $I_{c}$

\begin{figure}[!ht]
  \centering
    \includegraphics[width=0.2\textwidth]{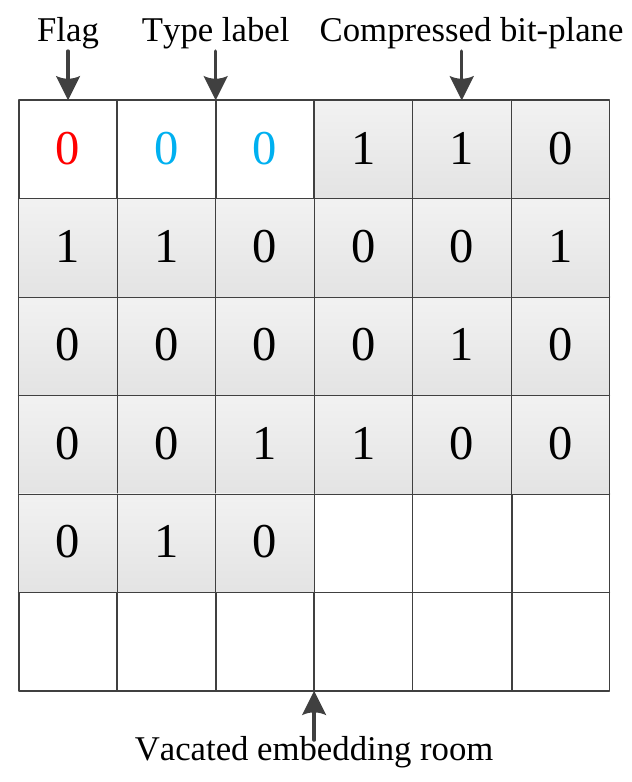}
  \caption{Example of a compressed bit-plane.}
\label{fig-7}
\end{figure}

\begin{figure*}[!ht]
  \centering
    \includegraphics[width=5in]{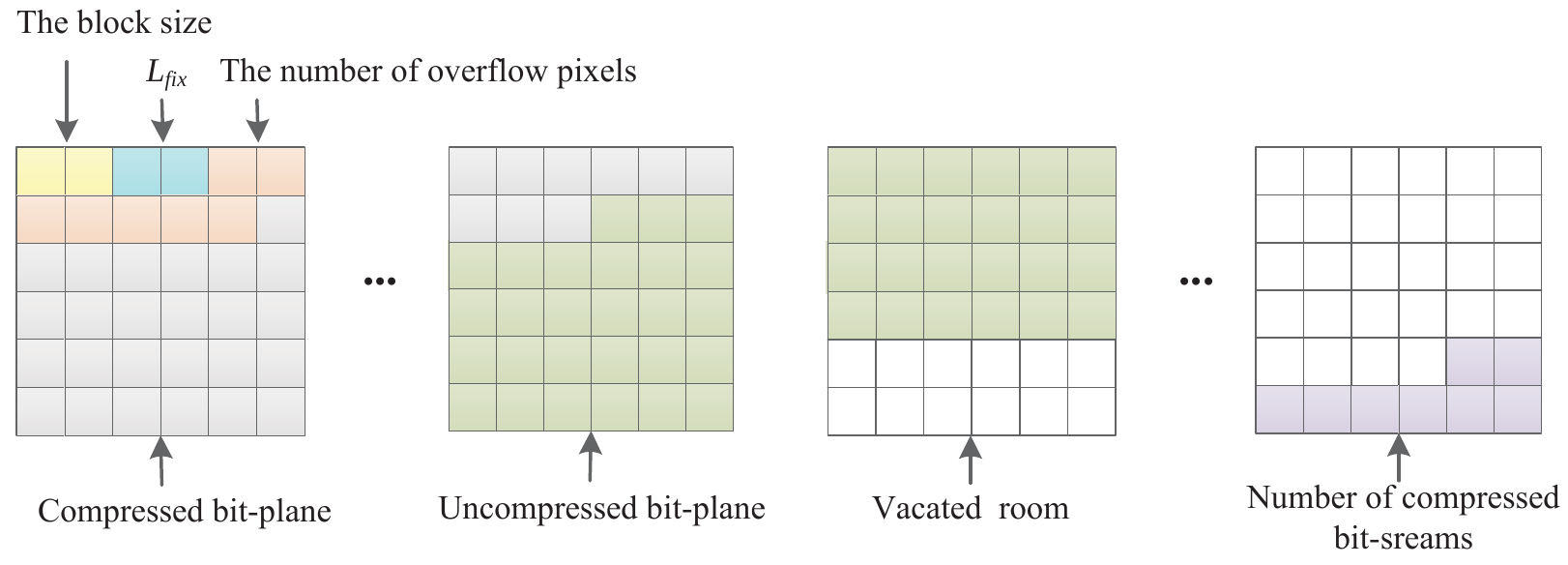}
  \caption{Example of vacating room.}
\label{fig-8}
\end{figure*}

\subsection{Image Encryption}
In the image encryption phase, generating a pseudo-random matrix $H$ of size $m\times n$ by key $K_{en}$. The pixel $x_{c}\left ( i,j \right )$ of the compressed image $I_{c}$  and the corresponding value $h\left ( i,j \right )$ of matrix $H$ are first converted into 8 bits binary sequence according to Eq. (5), respectively. Then, the bitwise exclusive-or (XOR) encryption is performed by using Eq. (7):
\begin{equation}
x_{e}^{k}\left ( i,j \right )=x_{c}^{k}\left ( i,j \right )\oplus h ^{k}\left ( i,j \right ),\: k=1,2...,8
\end{equation}
Where $x_{e}^{k}\left ( i,j \right )$  represents encrypted binary of the $k^{th}$ bit and $\bigoplus $ denotes XOR operation. Finally, the encrypted pixel  $x_{e}\left ( i,j \right )$ can be calculated according to Eq. (6). The encrypted image  $I_{e}$ is obtained by the above process.
\subsection{Data Hiding in Encrypted Image}
In this subsection, to embed the additional data in the vacated room, the number of compressed bit-streams placed on the LSB plane is first extracted and the location of the vacated room can be obtained according to the extracted information. Then, the data hiding key $K_{d}$ is adopted to encrypt the additional data. Finally, the encrypted additional data is embedded into the vacated multi-LSB planes by LSB substitution. As shown in Fig. 9, the location of the vacated room generated in Fig. 8 can be obtained according to the number of compressed bit-streams. Then, the marked encrypted image $I_{es}$  is generated by LSB substitution.
\begin{figure*}[!ht]
  \centering
    \includegraphics[width=5in]{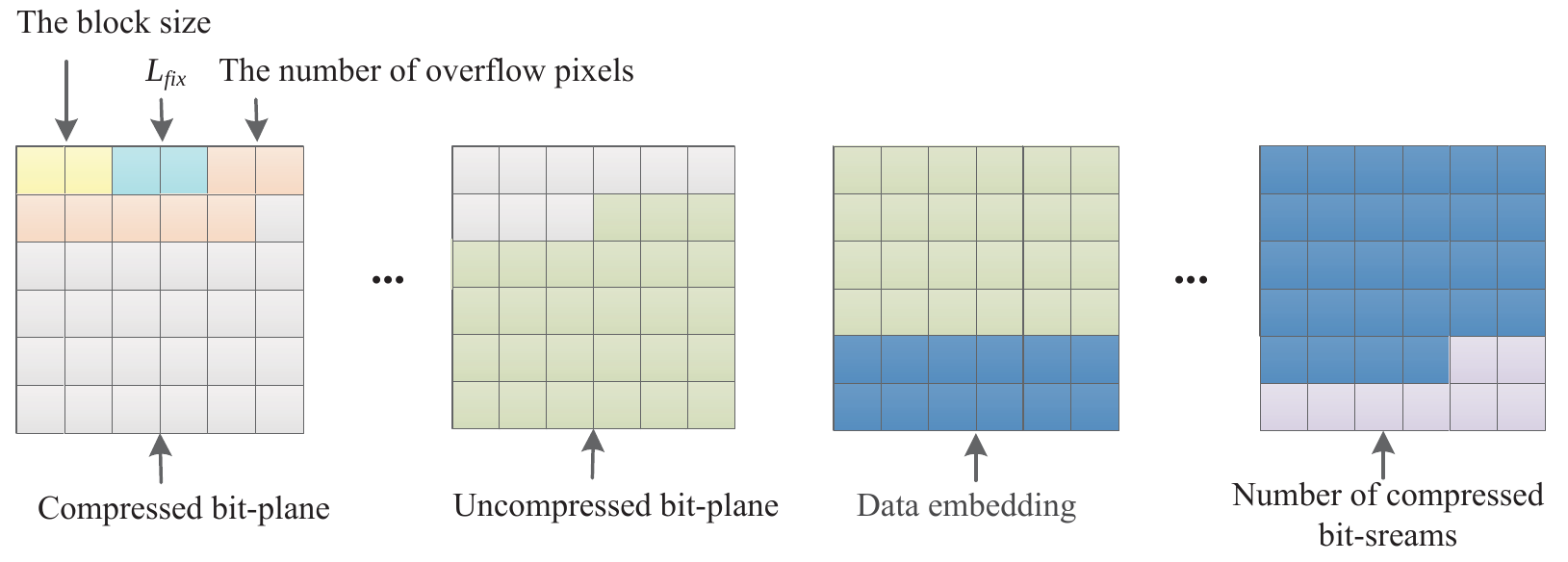}
  \caption{Example of embedding additional data.}
\label{fig-9}
\end{figure*}

\subsection{Data Extraction and Image Recovery}
The permission of data extraction and image recovery depend on the number of keys for the receiver. As shown in Fig. 10, there are three cases:
\par Case 1: The receiver only has the key $K_{d}$. Because the key $K_{en}$ only encrypts the image and does not affect the process of additional data, the extracted encrypted additional data can be correctly decrypted by key $K_{d}$. However, the original image cannot be recovered without the key $K_{en}$.
\par Case 2: The receiver only has the key $K_{en}$. The original image can be restored. Firstly, the block of size, $L_{fix}$ and the number of overflow pixel are extracted from MSB plane, and the number of compressed bit-streams for each bit-plane is extracted from LSB plane. Then, the image is decrypted by the key $K_{en}$  according to Eq. (7). The additional data cannot be extracted correctly without the key $K_{d}$.
\par Case 3: The receiver has both the key $K_{en}$ and $K_{d}$. The additional data can be extracted correctly and the original image can be recovered by the receiver.
\begin{figure*}[!ht]
  \centering
    \includegraphics[width=4in]{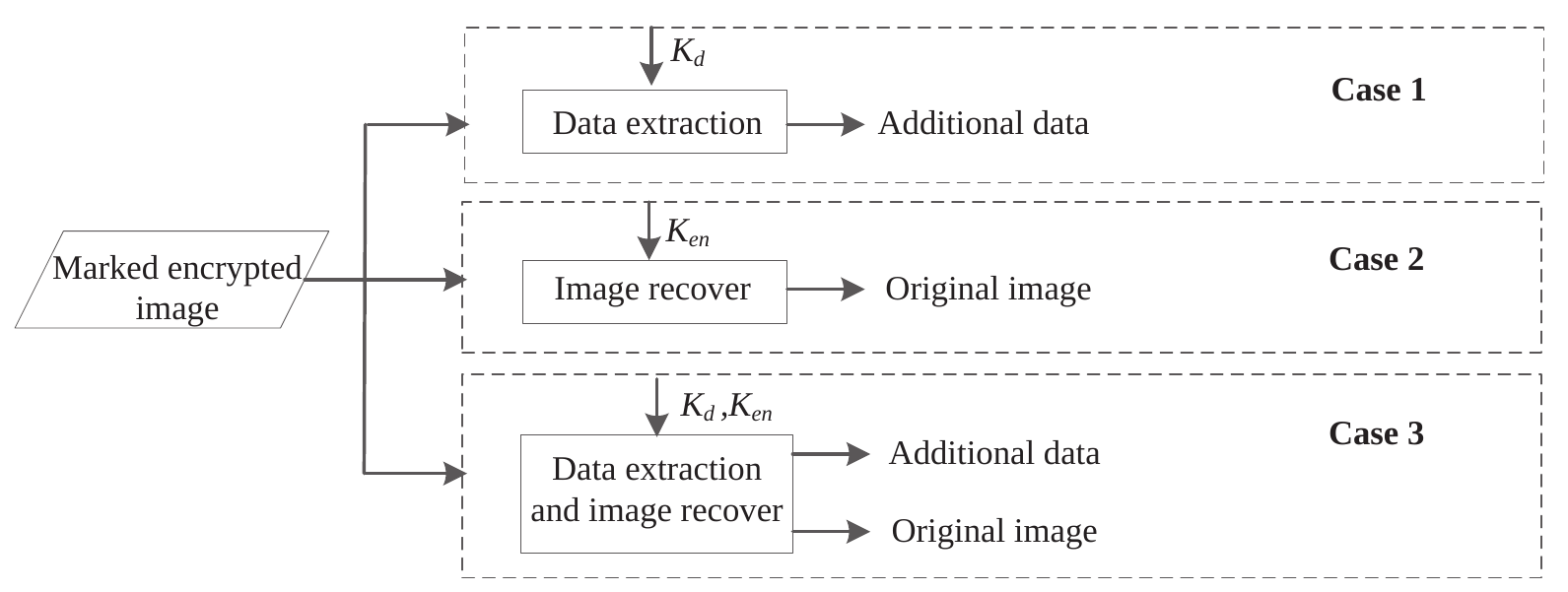}
  \caption{Three cases of data extraction and image recovery for receiver. }
\label{fig-10}
\end{figure*}

\section{Experimental Results and Discussion}
In this section, the reversibility and the net embedding capacity of the proposed method is first analyzed in Section IV-A. Then, in order to demonstrate that the proposed method outperforms the state-of-the-art methods, such as [19], [21][22], the experiments of embedding capacity is designed in Section IV-B. The peak signal-to-noise ratio (PSNR) and structural similarity (SSIM) are used to evaluate the reversibility of the method. At the same time, the embedding rate (ER) is used as the evaluation index of the embedding capacity. As shown in Fig. 11, six standard gray images: Airplane, Lena, Man, Jetplane, Baboon and Tiffany and three datasets: UCID [24], BOSSbase [25] and BOWS-2 [26] are used. Next, the experimental results and discussion will be introduced in detail.

 \begin{figure*}[!t]
\centering
\subfigure[]{\includegraphics[width=1in]{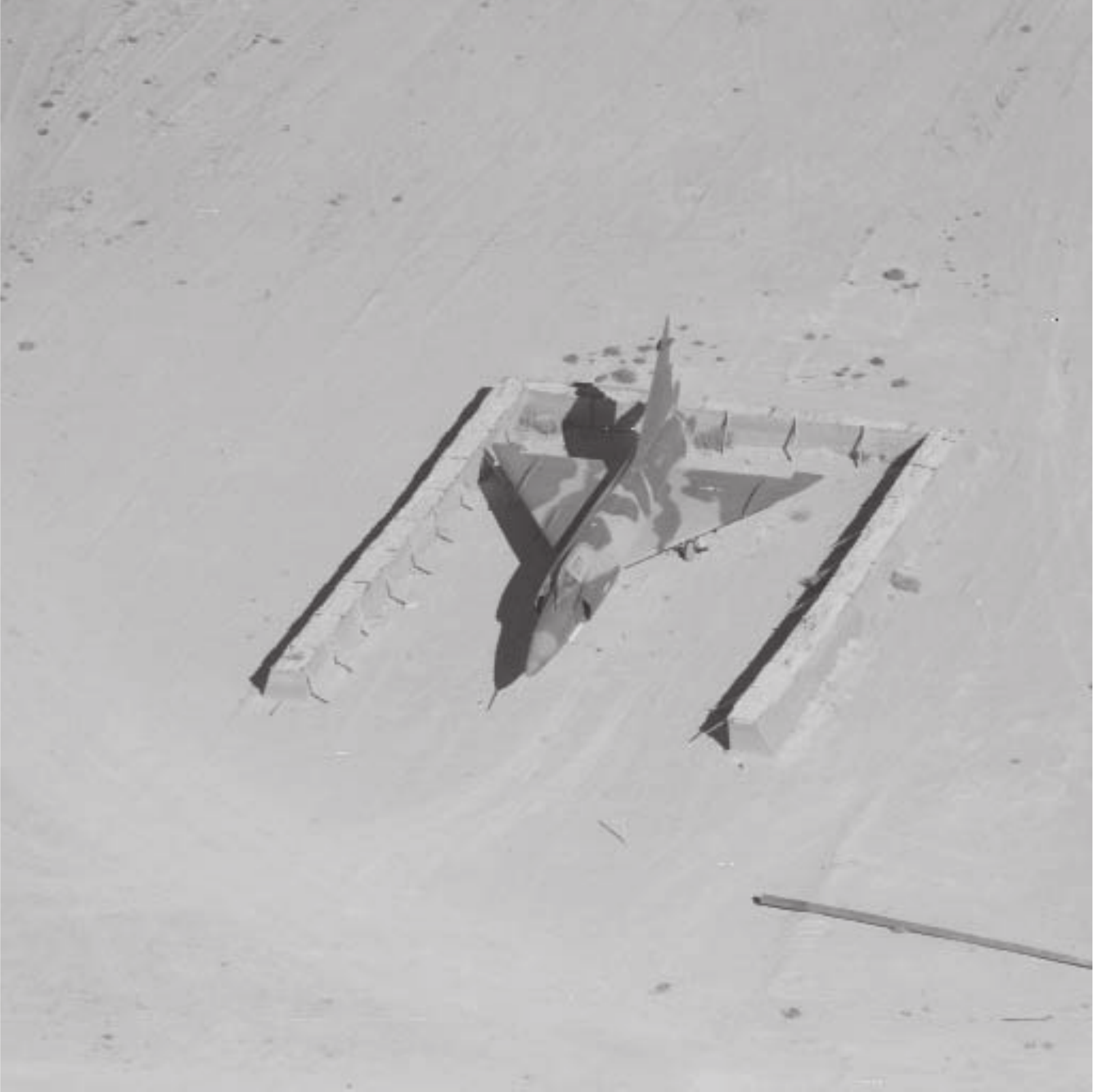}%
\label{fig-11-a}}
\hfil
\subfigure[]{\includegraphics[width=1in]{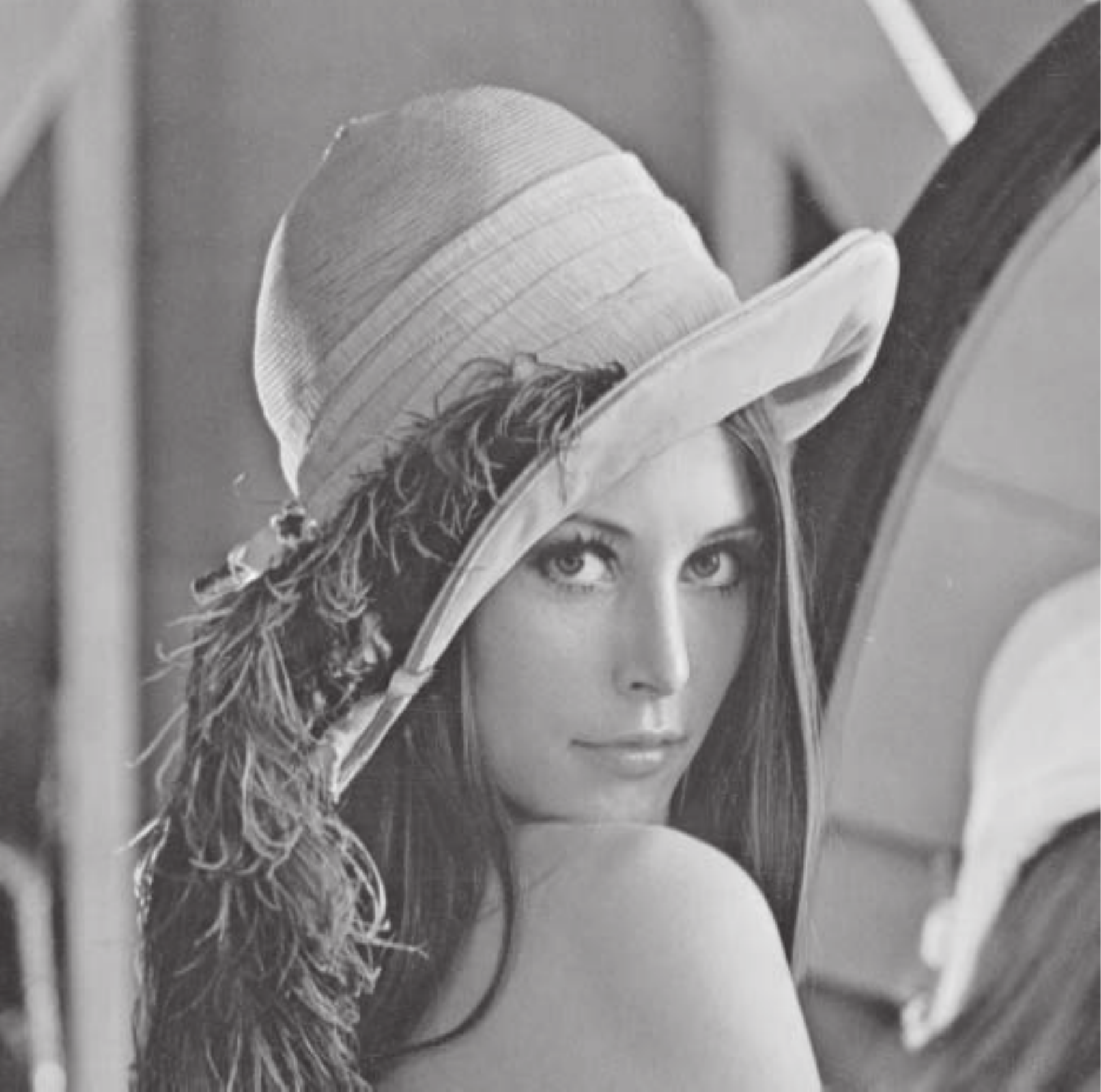}%
\label{fig-11-b}}
\subfigure[]{\includegraphics[width=1in]{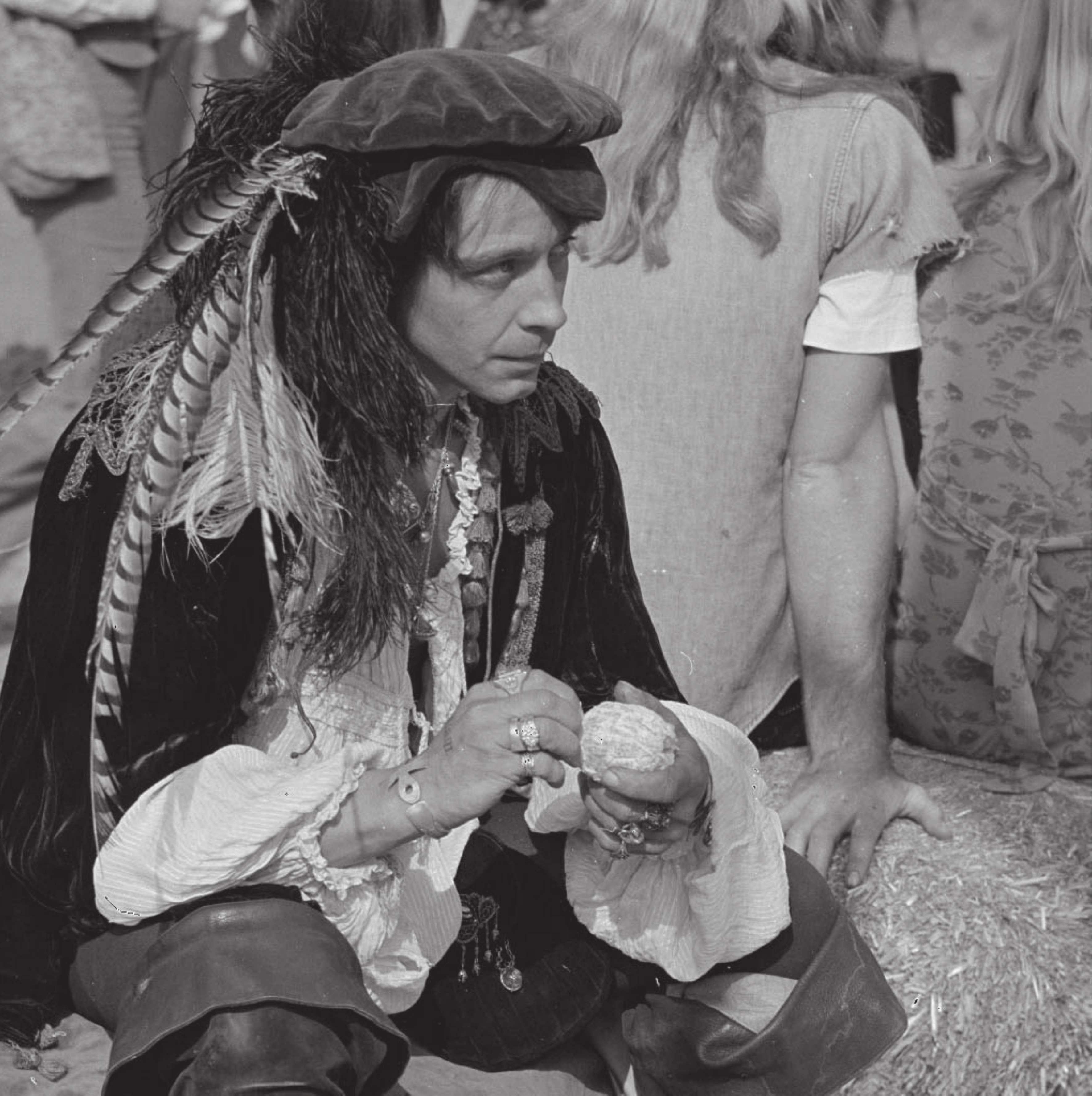}%
\label{fig-11-c}}
\hfil
\subfigure[]{\includegraphics[width=1in]{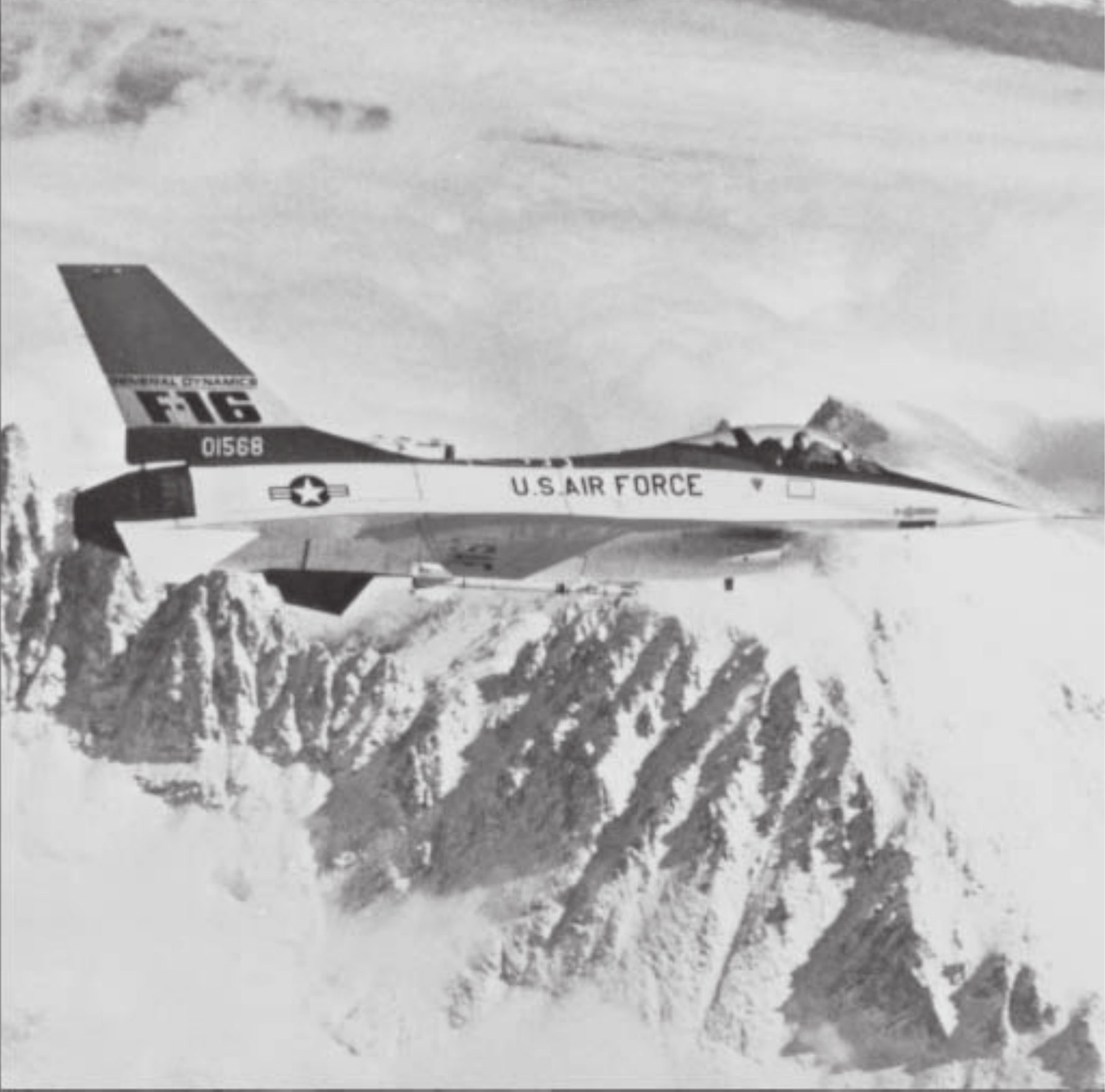}%
\label{fig-11-d}}
\hfil
\subfigure[]{\includegraphics[width=1in]{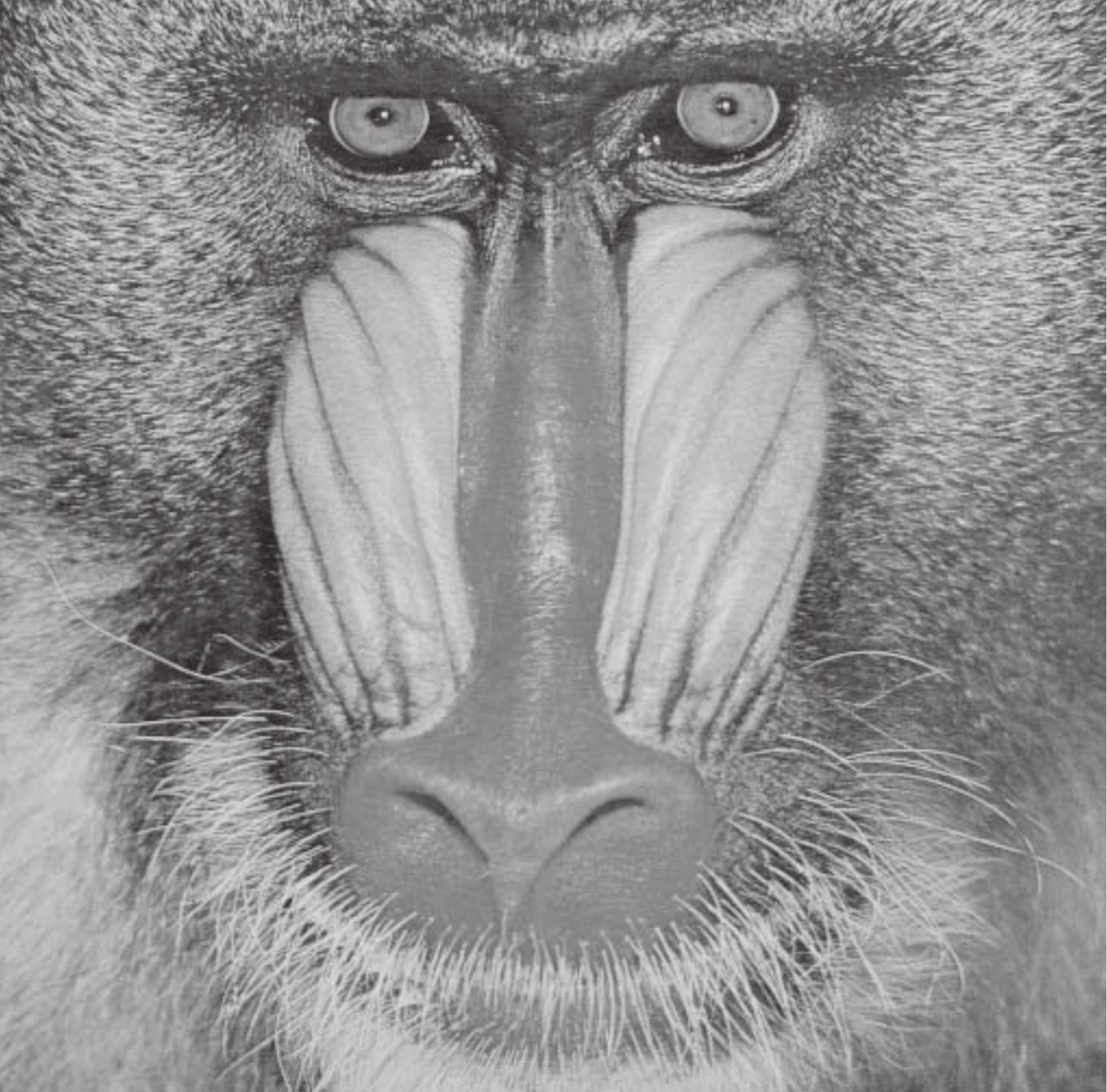}%
\label{fig-11-f}}
\hfil
\subfigure[]{\includegraphics[width=1in]{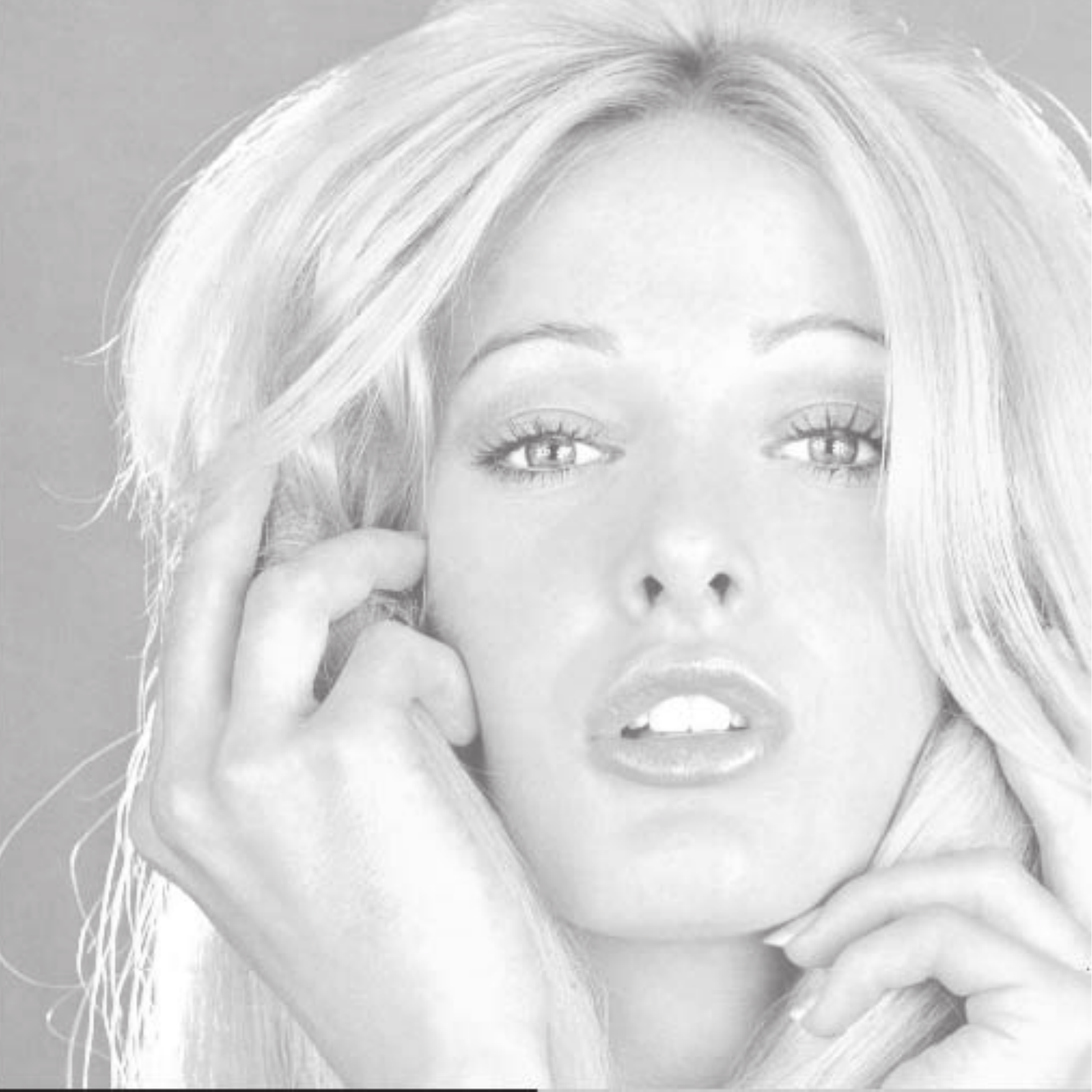}%
\label{fig-11-e}}
\caption{Two different frameworks of RDHEI methods: (a) VRAE and (b) RRBE.}
\label{fig-11}
\end{figure*}

\subsection{Performance Analysis of the Proposed Method}
\subsubsection{Discussion of Reversibility}
To demonstrate the original image of the proposed method can be reconstructed lossless, as shown in Fig. 12, Lena is taken as an example. Fig. 12(a) is the original image $I$, and Fig. 12(b) is the encrypted image $I_{e}$ obtained by the encryption key $K_{en}$. Fig. 12(c) shows the marked encrypted image $I_{es}$ that the ER is 3.0750 $bpp$. Since the data extraction and original image recovery can be separated in the proposed method, the content-owner can recover the original image using the encryption key $K_{en}$. Fig. 12(d) is the reconstructed image. The PSNR of the reconstructed image close to $+\infty$ and the SSIM equal to 1. That confirms the original image of the proposed method can be recovered lossless.

\begin{figure*}[!t]
\centering
\subfigure[]{\includegraphics[width=1in]{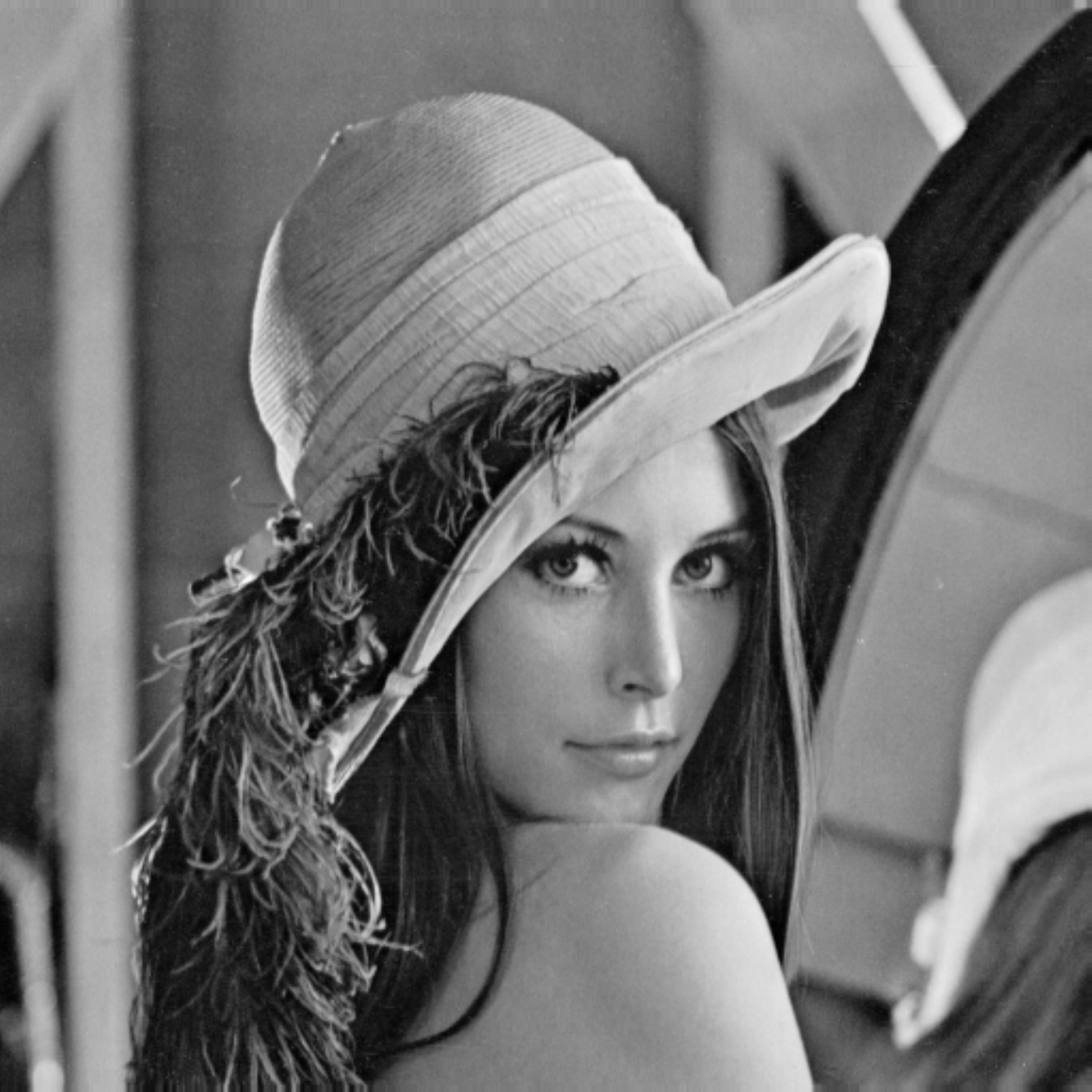}%
\label{fig-12-a}}
\hfil
\subfigure[]{\includegraphics[width=1in]{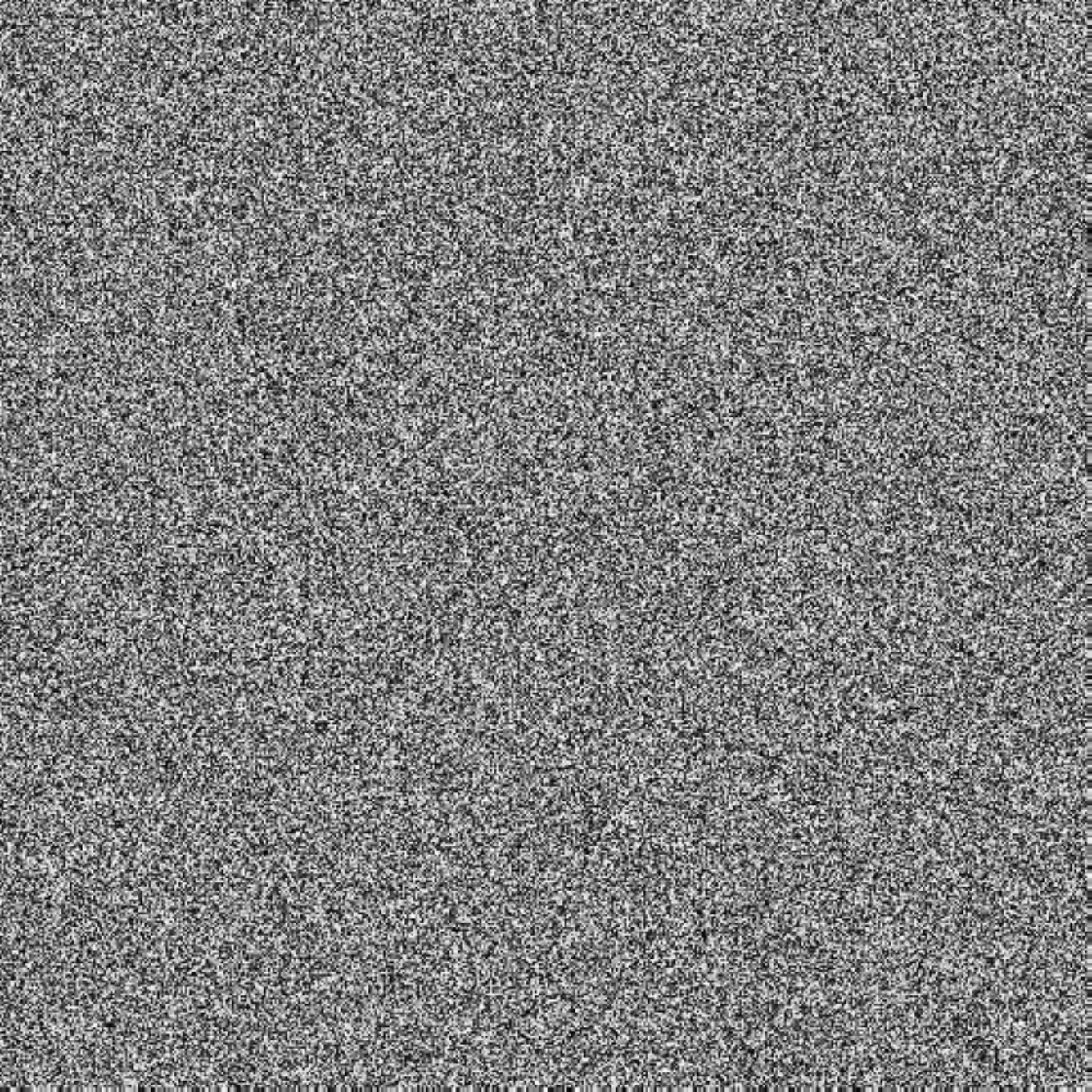}%
\label{fig-12-b}}
\subfigure[]{\includegraphics[width=1in]{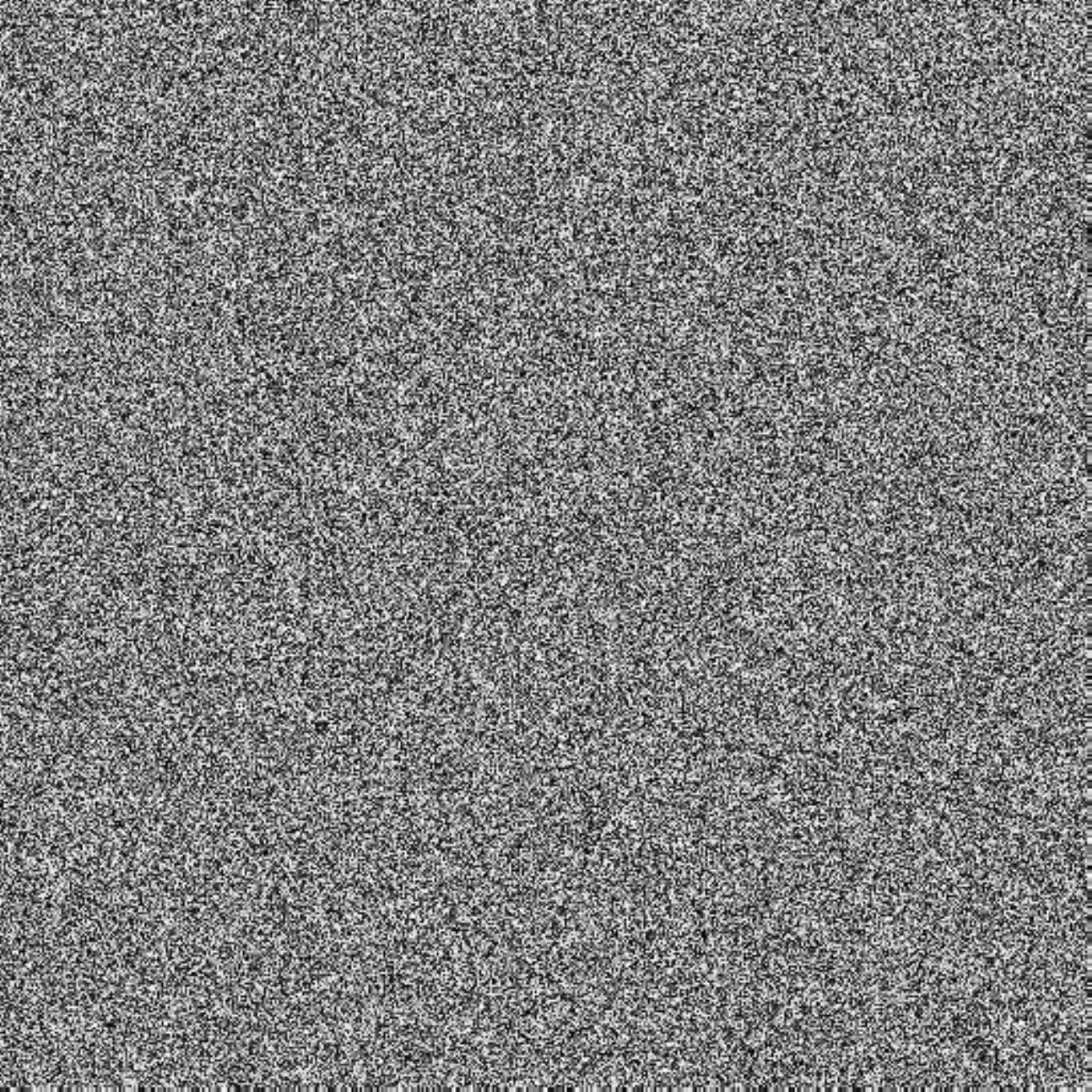}%
\label{fig-12-c}}
\hfil
\subfigure[]{\includegraphics[width=1in]{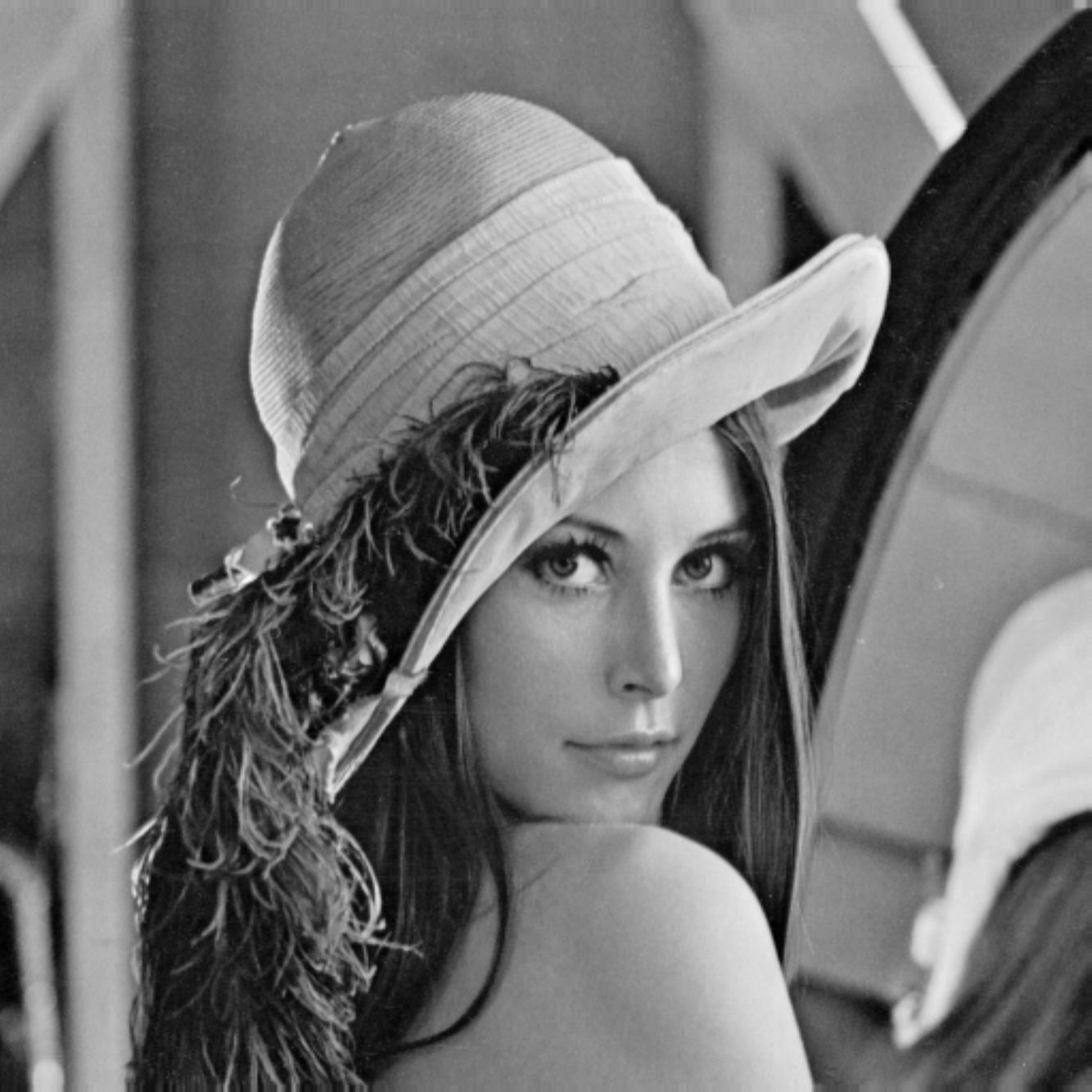}%
\label{fig-12-d}}
\caption{Two different frameworks of RDHEI methods: (a) VRAE and (b) RRBE.}
\label{fig-12}
\end{figure*}

\subsubsection{Discussion of Net Embedding Capacity}
The proposed is performed to evaluate net ER in the test images and three databases [24]-[26], respectively. Table I shows the number of bit-streams of each bit-plane before and after compression for six test images. On the one hand, since the $8^{th}$ bit-plane is used to record the sign of the prediction error, the number of ``0" and ``1" is similar and is randomly distributed in the bit-plane. Thus, the length of the compressed bit-streams is longer than that of the original bit-streams and the bit-plane is not compressed. On the other hand, because of the high accuracy of MED prediction, the number of ``0" is far greater than ``1" in the $7^{th}$-$5^{th}$ bit-planes and the length of the compressed bit-streams is less than that of the original bit-streams. Taken Airplane as an example, the length of compressed bit-streams is 5034 bits for the $7^{th}$ bit-plane. The compression degree is the largest compared with other bit-planes. The other five images have similar rules.
\par The length of auxiliary information, the net payload and ER are displayed in Table II on six test images. Assuming the number of overflow points is $t$ for an image, its number and location used 18 bits and $t\times 18$ bits to recorded, respectively. The block of size and the value of  $L_{fix}$ is denoted by 4 bits and 3 bits, respectively. The number of uncompressed bit-planes is recorded by $s$ bits. It can be observed from Table II that the length of auxiliary information is far shorter than the vacated room on six test images. Taken Airplane as an example, its auxiliary information is recorded by 118 bits, the net payload is 1005663 bits and the net ER is 3.8363 $bpp$. Table III displays the average net ER of the proposed method on three databases [24]-[26]. For BOSSbase database, the net ER of the best case is 7.8305 $bpp$, and the net ER of the worst case is 0.7416 $bpp$. For BOWS-2 database, the net ER of the best case is 7.0190 $bpp$, and the net ER of the worst case is 0.6239 $bpp$. But in UCID, the length of the auxiliary information of two images is longer than the vacated room. If these two images are not considered, the net ER of UCID the worst case is 0.2563 $bpp$.
\begin{table*}[!t]
\caption{The compressed bit-plane in different test images}
\label{table_1}
\centering
\begin{tabular}{|c|c|c|c|c|c|c|}
\hline
                            & Airplane                               & Lena                                   & Man                                    & Jetplane                               & Babooon                                & Tiffany                                \\ \cline{2-7}
\multirow{-2}{*}{Bit-plane} & 512*512                                & 512*512                                & 1024*1024                              & 512*512                                & 512*512                                & 512*512                                \\ \hline
Original bits             & 262144                                 & 262144                                 & 1048576                                & 262144                                 & 262144                                 & 262144                                 \\ \hline
8                           & -                                      & -                                      & -                                      & -                                      & -                                      & -                                      \\ \hline
7                           & 5034   &4443   & 17795  & 3564  & 39985  & 4375   \\ \hline
6                           & 17682  & 17933  &90490  & 20816  & 143101 & 20277  \\ \hline
5                           & 28408  & 58371  & 346108 & 57296  & 240610 & 59735  \\ \hline
4                           & 71013  & 161631 & 974860 & 116385 &  -               & 138435 \\ \hline
3                           &182697 & -                                      & -                                      & 220728 & -                                      & -                                      \\ \hline
2                           & -                                      & -                                      & -                                      & -                                      & -                                      & -                                      \\ \hline
1                           & -                                      & -                                      & -                                      & -                                      & -                                      & -                                      \\ \hline
\end{tabular}
\end{table*}

\begin{table*}[!t]
\caption{ The net payload and auxiliary information of different test images}
\label{table_2}
\centering
\begin{tabular}{|c|c|c|c|c|c|c|}
\hline
Test images   & Airplane                                & Lena                                   & Man                                     & Jetplane                               & Babooon                                & Tiffany                                \\ \hline
Room          & 1005781                                 & 806114                                 & 2764959                                 & 891826                                 & 362673                                 & 825670                                 \\ \hline
Number        & 5                                       & 0                                      & 125                                     & 1                                      & 3                                      & 5                                      \\ \hline
Location      & 18*5                                    & 18*0                                   & 18*125                                  & 18*1                                   & 18*3                                   & 18*5                                   \\ \hline
Parameter     & 10               & 11              & 11               & 10             &  12              &  10              \\ \hline
bits &  1005663 &  806085 &  2762428 & 891780 & 362589 &  825551 \\ \hline
\textit{bpp}  & 3.8363  & 3.0750 & 2.6345  & 3.4019 & 1.3832 & 3.1492 \\ \hline
\end{tabular}
\end{table*}

\begin{table*}[!t]
\caption{  Experimental results of the net ER ($bpp$) on three image databases}
\label{table_3}
\centering
\begin{tabular}{|c|c|c|c|}
\hline
Test images & UCID   & BOSSbase & BOWS-2 \\ \hline
Best case   & 5.2344 & 7.8305   & 7.0190 \\ \hline
Worst case  & 0.2563 & 0.7416   & 0.6239 \\ \hline
Average     & 2.8990 & 3.6248   & 3.4948 \\ \hline
PSNR        & +${ \infty }$      & +${ \infty }$        & +${ \infty }$      \\ \hline
SSIM        & 1      & 1        & 1  \\ \hline
\end{tabular}
\end{table*}

\begin{figure*}[!ht]
  \centering
    \includegraphics[width=5in]{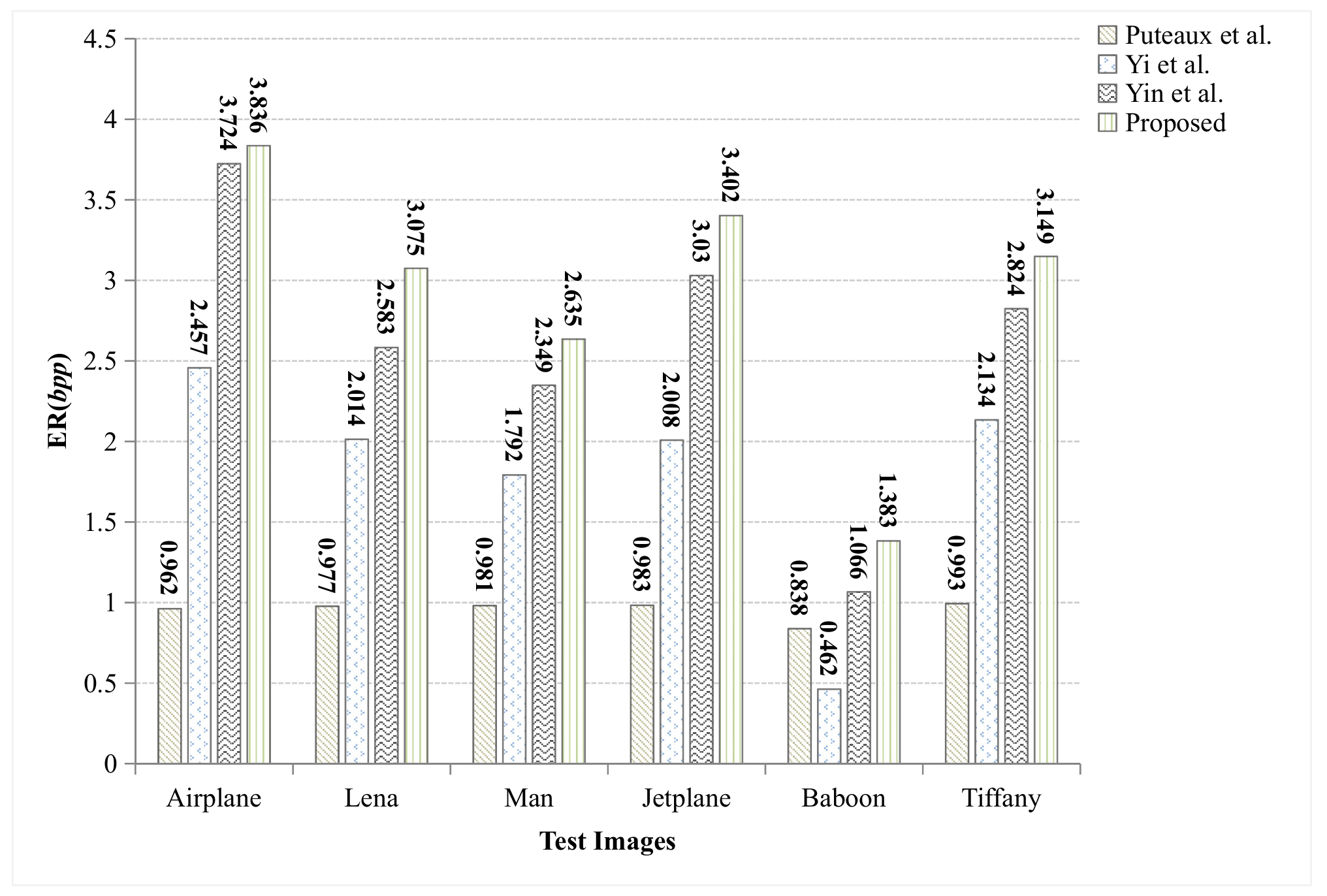}
  \caption{Comparison of ER ($bpp$) on six test images.}
\label{fig-13}
\end{figure*}

\begin{figure*}[!ht]
  \centering
    \includegraphics[width=5in]{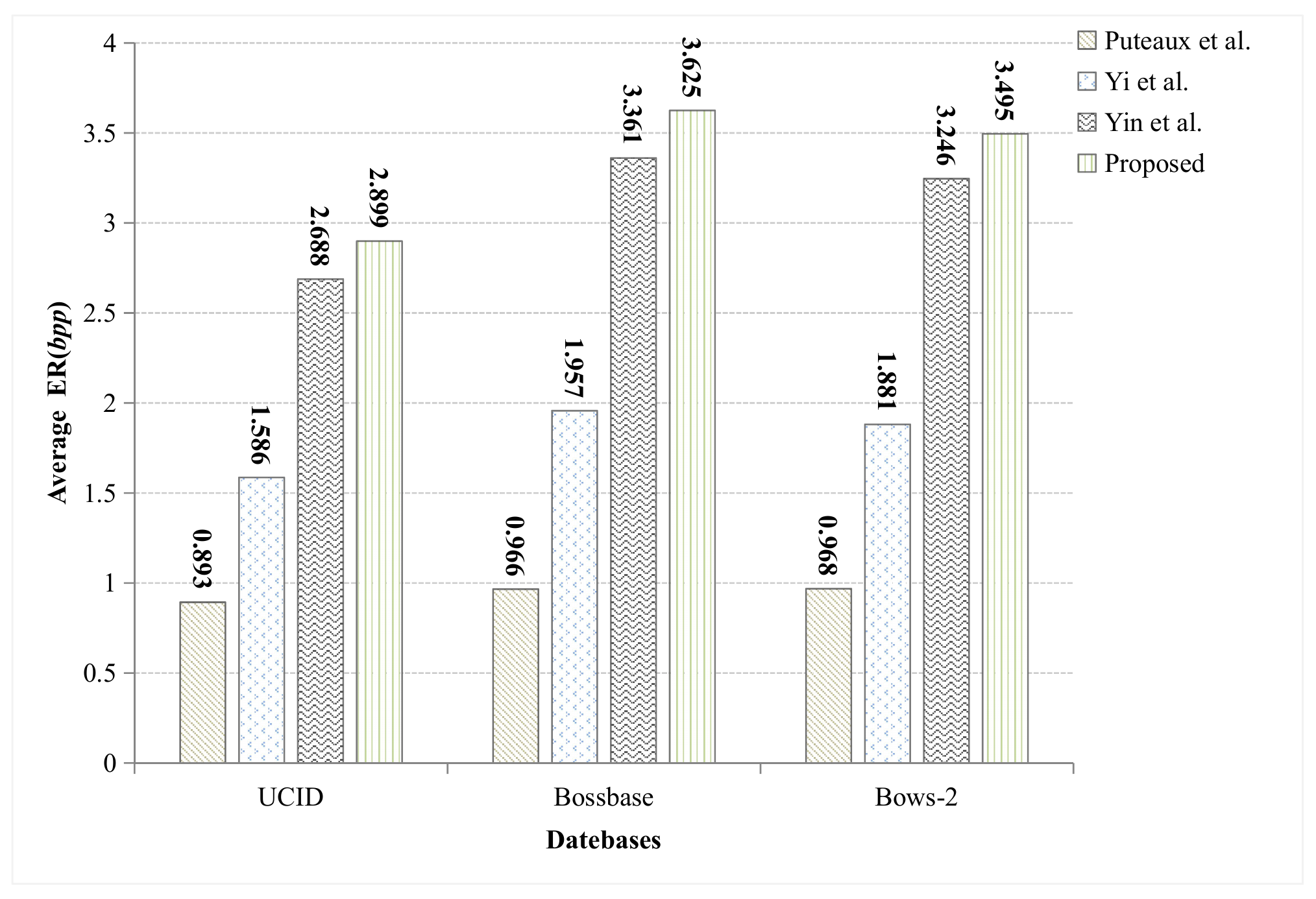}
  \caption{Comparison of average ER ($bpp$)  on three image databases [24]-[26].}
\label{fig-14}
\end{figure*}

\subsection{Performance Comparison}
Since the proposed method and the state-of-the-art methods, such as [21][22], all can reconstruct the original image lossless, the part only gives the performance comparison in terms of ER.
\par Fig. 13 shows the experimental results of the ER on six test images. Compared with the methods proposed by Puteaux ${ et~al. }$ [19], Yi ${ et~al. }$ [21] and Yin ${ et~al. }$ [22], the experimental results confirm that the ER of the proposed method is the best. Taken Lena as an example, the ER of the proposed method is 3.0750 $bpp$, while Puteaux ${ et~al. }$ [19], Yi ${ et~al. }$ [21] and Yin ${ et~al. }$ [22] methods are 0.977, 2.014 and 2.583 $bpp$, respectively. In order to further demonstrate that the ER of the proposed method outperforms the state-of-the-art methods, Fig. 14 displays the comparative experiments on the three databases [24]-[26]. Since the MSB of the available pixels is substituted by 1 bit additional data in Puteaux ${ et~al. }$ [19], the average ER is less than 1 $bpp$. Yin ${ et~al. }$ [22] use the multi-MSB substitution to improve the average ER. In the UCID [24], the average ER of the proposed method is a significant improvement compared with the methods proposed by Puteaux ${ et~al. }$ [19], Yi ${ et~al. }$ [21] and Yin ${ et~al. }$ [22]. The average ER of the proposed method is 2.899 $bpp$, while Puteaux ${ et~al. }$ [19], Yi ${ et~al. }$ [21] and Yin ${ et~al. }$ [22] methods are 0.893, 1.586 and 2.688 $bpp$, respectively. There are similar rules in the BOSSbase [25] and BOWS-2 [26].

\section{Conclusions}
In this paper, an RDHEI scheme based on pixel prediction and bit-plane compression is proposed. since the bit-planes of prediction error have a large number of adjacent ``0" or ``1",  the characteristic is utilized to vacate more room. Experiments show that not only the image can be recovered and the embedded data can be extracted for the receiver, separately,  but also the embedding capacity of the proposed method not only outperforms the state-of-the-art methods.
\par In future work, the embedding capacity will be enhanced in two ways. On the one hand, we inverse the part of the MSB and LSB and choose the shortest compression length. On the other hand, two different compression algorithms can be utilized for multi-MSB planes and multi-LSB planes. The proposed method can explore higher data embedding capacity.


%



\section*{Acknowledgments}

This research work is partly supported by National Natural Science Foundation of China (61872003, U1636206, 61860206004).

\ifCLASSOPTIONcaptionsoff
  \newpage
\fi

\end{document}